\documentclass[prd,aps,12pt,preprintnumbers,amsmath,amsssymb,epsf,psfig,nofootinbib]{revtex4}

\def\be{\begin{equation}}
\def\ee{\end{equation}}
\def\bea{\begin{eqnarray}}
\def\eea{\end{eqnarray}}

\def\ssc{\scriptscriptstyle}
\def\lsim{\mathrel{\raise.3ex\hbox{$<$\kern-.75em\lower1ex\hbox{$\sim$}}} }
\def\gsim{\mathrel{\raise.3ex\hbox{$>$\kern-.75em\lower1ex\hbox{$\sim$}}} }

\begin{document}
\draft
\preprint{{\vbox{\hbox{NCU-HEP-k023} \hbox{Jan 2006}\hbox{ed. Jun
2006} }}}

\vspace*{1.5in}
\title{Physics of Quantum Relativity through a Linear Realization}

\author{\bf Ashok Das$^1$ and Otto C. W. Kong$^2$\vspace*{.2in}}
\email{otto@phy.ncu.edu.tw}

\affiliation
{ $^1$Department of Physics, University of Rochester, Rochester, NY 14627, USA \\
$^2$Department of Physics, National Central University, Chung-li, Taiwan 32054}

\begin{abstract}
The idea of  quantum relativity as a generalized, or rather
deformed, version of Einstein (special) relativity has been taking
shape in recent years. Following the perspective of deformations,
while staying within the framework of Lie algebra, we implement
explicitly a simple linear realization of the relativity symmetry,
and explore systematically the resulting physical interpretations.
Some suggestions we make may sound radical, but are  arguably
natural within the context of our formulation. Our work may provide
a new perspective on the subject matter, complementary to the
previous approach(es), and may lead to a better understanding of the
physics.
\end{abstract}
\maketitle

\newpage
\section{Introduction}
In recent years, a new form of (special) relativity has been
introduced under the names deformed special relativity or doubly (or
triply) special relativity (DSR, TSR) \cite{dsr,dsr2,tsr}. It is
motivated from a desire to incorporate additional invariant
(dimensional) parameters into the theory {beyond the speed of
light}, particularly an invariant quantum scale such as the Planck
scale. (This idea  {actually} dates back to a paper by Snyder
\cite{S} which is also a precursor to the idea of non-commutative
geometry.) As a result, the new  relativity can really be thought of
as the quantum relativity. Relativity, of course, involves the
behavior of frames of reference in physics and the relativity
algebra is the algebra of transformations of the reference frames
which also reflects the algebra of ``space-time symmetry".
Therefore, one can view the algebra of quantum relativity as the
algebra of transformations of quantum reference frames or the
symmetry algebra of quantum space-time. We note that some discussion
about quantum frames of reference already exists in literature and
we want to refer the reader particularly to Refs.\cite{AK,R}. The
first of the two references discusses such issues within the context
of non-relativistic quantum mechanics while the second concerns
theories under the influence of gravity. The shared conclusion of
the two papers is that a quantum frame of reference has to be
characterized by its mass. In the case of gravity, it is illustrated
that the gravitational properties of the reference frame itself need
to be taken into account in order to define local gauge invariant
observable. On the other hand, there is also the very intriguing
notion of a quantum space-time structure. Space-time structure
beyond a certain microscopic scale certainly cannot be physically or
operationally defined as the commutative geometry of (Einstein)
special relativity. One hopes that understanding the (special)
quantum relativity would pave the way to understanding the quantum
structure of space-time, and eventually even a quantum theory of
gravity may be constructed as the general theory of quantum
relativity. The present article is an attempt in this direction.

Even when one knows the mathematical description of the
transformations, the physical interpretation may be much harder to
come by. As we know, while Lorentz had already studied the (Lorentz)
transformations, it was Einstein who {brought out} the correct
physical meaning {of} these transformations\cite{Mil}.  In trying to
obtain the correct physical picture from a mathematical description,
one has to have an open mind for unconventional perspectives (that
may arise) on some of the most basic notions about physics, in
general, and space-time structure, in particular.  Most of the
discussions of the (new) relativity so far have  focused on
nonlinear realizations of the symmetry algebra that arises. There is
still no agreement among different authors on what the ultimate
algebra of quantum relativity {is}. Here we will focus on {\small
\boldmath $SO(1,5)$} as the ultimate Lie algebra of the symmetry of
quantum relativity  which has essentially been advocated within the
context of triply special relativity (TSR)  \cite{tsr}, although
most of our discussion on doubly special relativity (DSR) \cite{dsr}
as an immediate structure will still be valid if that corresponds to
quantum relativity, namely, if the intermediate structure in our
discussions coincides with the final. From the point of view of Lie
algebra deformation,  {\small \boldmath $SO(1,5)$} has actually been
identified essentially {with} the natural {stabilizer} of the
``Poincar\'e + Heisenberg" symmetry \cite{spha}. Physical
observations with limited precision can never truly confirm an
unstable algebra as the symmetry. Correspondingly, this perspective
is strongly suggestive of taking {\small \boldmath $SO(1,5)$} as the
natural candidate for the symmetry algebra of quantum relativity. It
is also very reassuring that this symmetry {algebra} arises
naturally from the perspective of quantum relativity as a
deformation {of} special relativity  \cite{tsr,CO}.

We take the structure of the Lie algebra seriously as {denoting} the
symmetry of ``space-time" and focus on a linear realization with a
classical or commutative geometry as the background of quantum
relativity. Such a linear realization has been discussed to some
extent in the literature, but neither systematically nor in detail.
Our approach here is to try to understand the true physical
implications of  quantum relativity directly from a study of the
transformations of the  quantum reference frames and try to deduce
the {underlying} geometric structure as an extension of the
conventional space-time. We consider such an analysis as {being}
complementary to the earlier studies involving nonlinear
realizations. We note that our approach is largely inspired by
Ref.\cite{GL} which, in our opinion,  has brought up {some}
interesting perspectives related to linear realization without
putting all of them on more solid foundation. In trying to do this,
we arrive at some very interesting and unexpected results. The most
interesting among them is the extension of Einstein space-time
structure into a higher dimensional geometry which is {\rm not} to
be interpreted as an extended space-time in the usual sense. This
result is unconventional, but is of central importance to our
discussions. We obtain the symmetry of quantum relativity through
the approach of  deformations and look for direct implications. We
ask for sensible interpretations of mathematical results, and make
suggestions along the {way}. Our analysis should be thought of as an
initial attempt, rather than a final understanding.  We believe that
there are still a lot more questions to be understood than the ones
we discuss with suggested possible answers. We have put forward
ideas here which seem to fit the physical problem at hand. Some of
this is unconventional, but we think that they are quite reasonable
and plausible and are in the right direction.

The article is organized as follows. In the next section, we write
down explicitly the two-step deformation procedure to arrive at the
quantum relativity, more or less following Ref.\cite{GL}. In
Sec.III, we focus only on the first deformation introducing the
invariant ultraviolet scale (the Plank scale), which gives a DSR
structure. Here, the linear realization necessitates the
introduction of a new geometric dimension along with the 4D
space-time. The corresponding new coordinate has {the canonical}
dimension of time over mass while having a spacelike geometric
signature. It also suggests a new definition of energy-momentum as
{a} coordinate derivative in the ``nonrelativistic limit".  This
section has the most dramatic or radical results {and all of this}
fits in well with the notion of quantum frames of reference. In
Sec.IV, we discuss relations to noncommutative or quantum (operator)
realization of 4D space-time. In Sec.V, we focus on the geometric
structure of the last deformation introducing an invariant infrared
scale (the cosmological constant). {Subsequently} we address some
important issues about the quantum relativistic momentum in Sec. VI
before we conclude the article in the last section.

\section{Quantum Relativity through Deformations}
Let us start by writing down the Lie algebra of {\small \boldmath
$SO(m,n)$} (with signature convention for the metric starting with a
+)
\be \label{so}
[J_{\ssc A\!B}, J_{\ssc L\!N}] =  i\, ( \eta_{\ssc
B\!L}  J_{\ssc A\!N} - \eta_{\ssc A\!L}  J_{\ssc B\!N} + \eta_{\ssc
A\!N}  J_{\ssc B\!L} -\eta_{\ssc B\!N}  J_{\ssc A\!L}) \;,
\ee where
indices $A,B,L,N$ take values from $0$ to $d-1$. For $d=4$, we have
the familiar algebra {\small \boldmath $SO(1,3)$}  describing
special relativity. However, we would like to start our discussion
with  {\small \boldmath$SO(0,3)$}  (which coincides with {\small
\boldmath $SO(3)$})  as a relativity algebra.  As we know, Newtonian
physics is described on {a} three-dimensional space. {The}  symmetry
algebra for the rotational invariance represented by {\small
\boldmath $SO(3)$} {with the corresponding generators given by}
\be
M_{ij} = i \, (x_i \, \partial_j - x_j \, \partial_i ), \quad
i,j=1,2,3 \;.
\ee
In this case, there is no index 0 and the metric
has the (special) signature $\eta_{ij}= (-1,-1,-1)$. We  have the
coordinate representation of the 3-momentum given by
 $p_i=  i\hbar  \, \partial_i =   {i\hbar  \,   \frac{\partial}{\partial x^i}}$
 (we take $\hbar=1$ in the following). The rotations can be augmented by {the}
three-dimensional translations to {define} the complete symmetry
group of 3D space. An arbitrary symmetry transformation can be taken
as a transformation between two (inertial) frames of reference. In
this case there is an alternative special way of getting a
translation, namely,
\be
x^i \to x^i + \Delta x^i\;, \quad \Delta
x^i(t) = v^i  t\;,
\label{translation} \ee
where $t$ denotes a
parameter outside of the three-dimensional manifold with $v^{i}$
given by $\frac{dx^{i}}{dt}$, the velocity. The parameter $t$ is
identified with the (absolute) {\em time} and such translations are
known as Galilean boosts. To the extent that  {\em time} is just an
external parameter, the Galilean boosts do not distinguish
themselves from translations. The relevant symmetry group describing
the admissible transformations between reference frames is the group
{\small\boldmath $ISO(3) \equiv SO(3) \otimes_{s} R^3 $} where
{\small \boldmath $\otimes_{s}$} represents the semi-direct product.
The generators of Galilean boosts (or special translations) can be
denoted by $N_i$ and satisfy \be \label{bst} [M_{ij}, N_k] = i\,
(\eta_{jk} N_i - \eta_{ik} N_j) \;. \ee Note that much like the
momentum, we can have the coordinate representation $N_i = i \,
\partial_i$ which satisfies Eq.(\ref{bst}). Of course, $N_i$'s
commute among themselves and we have Galilean relativity.

Within the framework of Galilean relativity, the speed of a particle
(as well as the speed of inertial reference frames) can take any
value. Einstein realized that one has to go beyond Galilean
relativity in order to accommodate an invariant speed of light, $c$.
From the present perspective, one can extend the three dimensional
manifold to a four dimensional one by including (the external
parameter) $t$ so that $x^{\mu} = (x^{0}, x^{i}) = (c \, t, x^{i}),
\mu = 0,1,2,3$. Furthermore, if one introduces the velocity
four-vector on this manifold as $u^{\mu} = (u^{0}, u^{i})$ with $u^0
= c/\sqrt{c^2-v^2} =\gamma,  u^i= v^i/\sqrt{c^2-v^2} = \gamma
\beta^i$ and $\beta^i = v^i/c$ ($c$ represents the speed of light)
{so} that
\be
\eta_{\mu\nu} u^{\mu}u^{\nu} = (u^0)^{2} - (u^i)^{2} =
1 \; ,\label{4}
\ee
then Eq.(\ref{4}) equivalently leads to
\be
-
\eta_{ij} v^i v^j = v^2 = c^2 \left(1-\frac{1}{\gamma^2}\right) \leq
c^2,\label{5}
\ee
with equality attained only in the limit $\gamma
\to \infty$. With {this} constraint  the velocity (speed) takes
values on the coset space  $v^i \in${\small \boldmath
$SO(1,3)/SO(3)$} in contrast to the case of Galilean relativity
where $v^i \in${\small\boldmath $R^3 \equiv  ISO(3)/SO(3)$}.
Furthermore, extending the manifold to a four dimensional one, we
obtain a linear realization of the transformation group of reference
frames which is deformed to {\small \boldmath $SO(1,3)$}, the
Lorentz group. The deformation of the algebra is given by
\be [N_i,
N_j]  \longrightarrow -i\, M_{ij} \; ,
\ee
and the $N_i$'s can now
be identified with {the} $M_{{\ssc 0}i}$'s in this extended
``space", {the} space-time {manifold}, so that the full set of six
$M_{\mu\nu} (= J_{\mu\nu}$) satisfying Eq.(\ref{so}) can be written
as
\be \label{lor}
M_{\mu\nu} = i (x_\mu \partial_\nu -x_\nu \,
\partial_\mu) \; .
\ee
Furthermore, adding the four-dimensional translations with
$p_{\ssc 0} = E/c = i\, \partial_{\ssc 0} = i\,
\frac{\partial}{\partial x^{\ssc 0}}$ ($\hbar=1$), we obtain the
full symmetry for Einstein special relativity described by the
Poincar\'e  group,  {\small\boldmath $ISO(1,3) \equiv SO(1,3)
\otimes_{s} R^4 $} which represents the complete transformation
group of inertial reference frames.

The idea of quantum relativity is to introduce  further invariant(s)
into the  physical system. In special relativity, for example, we
note that the momentum four vector can take any value. An invariant
bound for the  four-momentum (energy-momentum four-vector) that
serves as a bound for elementary quantum states is one such example
and such a generalization is commonly {referred to as}  ``doubly
special relativity" (DSR)  \cite{dsr}. One can follow the example of
Einstein relativity and derive this generalization as follows. Let
us consider a parameter $\sigma$ outside of the four-dimensional
space-time manifold and consider special translations in this
manifold of the form (similar to Galilean boosts, see
Eq.(\ref{translation}))
 \be
 x^{\mu}\to x^{\mu} + \Delta x^{\mu}\;,\quad \Delta x^{\mu} (\sigma)
= V^{\mu} \sigma\;,\label{vmu}
 \ee
where we have identified $V^{\mu} = \frac{dx^{\mu}}{d\sigma}$. The
generators $O_{\!\mu}$'s of these translations have the same
commutation relations as the conventional  four-dimensional
translations. Furthermore, let us extend the 4D space-time manifold
by adding the coordinate $\sigma$ as $x^{\ssc A} = (x^{\mu}, x^{4})
= (x^{\mu}, \kappa c\,\sigma), A=0,1,2,3,4$, where $\kappa$ denotes
the Planck mass and $c$ the speed of light. With this particular
choice of the extra coordinate, we recognize that $\sigma$ has the
dimensions of time/mass. As a result $V^{\mu}$ in Eq.(\ref{vmu}) has
the dimension of a momentum and we identify $V^{\mu} = p^{\mu}$. (We
will discuss more on this identification in the next section.) It is
clear that like the velocity in the case of Galilean relativity,
here $p^{\mu}$ can take any value. However, following the discussion
in the case of Einstein relativity, we see that on this
five-dimensional manifold, if we define a momentum 5-vector
$\pi^{\ssc A}$ as
\footnote{We scale by a factor $\kappa$ relative
to the common notation as first introduced by Snyder \cite{S}.}
\be
\label{pi}
\pi^{\ssc A} = \left(\pi^{\mu}, \pi^{\ssc 4}\right) =
\left(\frac{p^{\mu}}{\sqrt{\kappa^{2}c^{2} - p_{\mu}p^{\mu}}},
\frac{\kappa c}{\sqrt{\kappa^{2}c^{2} - p_{\mu}p^{\mu}}}\right) =
\left(\Gamma  \alpha^\mu  ,\Gamma \right) \; ,
\ee
with
$\alpha^{\mu} = p^{\mu}/\kappa c$, this will satisfy
\be \label{pi2}
\eta_{\ssc A\!B} \pi^{\ssc A} \pi^{\ssc B}  = \eta_{\mu\nu}
\pi^{\mu}\pi^{\nu} - \left(\pi^{\ssc 4}\right)^{2} = -1 \; .
\ee
This, in turn, would imply {that} (see Eq.(\ref{5}))
\be
\label{Gamma}
p_{\mu}p^{\mu} = \eta_{\mu\nu} p^\mu p^\nu = \kappa^2
c^2 \left(1-\frac{1}{\Gamma^2}\right) \leq \kappa^2 c^2 \;,
\ee
where as we have said earlier $\kappa$ stands for the Planck mass.
Similar to the earlier discussion on Einstein relativity,  we see
{that in this case} the four momentum lives on the coset space
$p^\mu \in ${\small \boldmath $SO(1,4)/SO(1,3)$} instead of $p^\mu
\in ${\small \boldmath $R^4$} and the de Sitter group {\small
\boldmath $SO(1,4)$} corresponds to the symmetry of the deformed
relativity here.

In this extended (five-dimensional) manifold, the extra generators
needed to complete {\small \boldmath $SO(1,4)$} and {to} lead to a
linear realization can now be taken as
\be
O_{\!\mu}  \equiv
J_{\mu\ssc 4} = i\, (x_\mu \partial_{\ssc 4} -x_{\ssc 4}\,
\partial_\mu) \;.\label{O}
\ee
Like the {conventional 4D} translation generators, the
generators $O_{\!\mu}$'s also satisfy (we note the identification
made earlier $M_{\mu\!\nu} = J_{\mu\!\nu}$)
\be [M_{\mu\!\nu},
O_{\!\lambda}] =  i \,(\eta_{\nu\!\lambda} O_{\!\mu} -
\eta_{\mu\!\lambda} O_{\!\nu}) \;. \ee
However, with the
identification in Eq.(\ref{O}),  the algebra is deformed to {\small
\boldmath $SO(1,4)$} with
\be [O_{\!\mu}, O_{\!\nu}] \longrightarrow
i\, M_{\mu\nu} \;.\label{O1}
\ee
We call the transformations
generated by $O_{\!\mu}$'s as de Sitter momentum boosts, or simply
as momentum boosts. {Adding the five-dimensional translations, the
full symmetry group of this manifold becomes {\small\boldmath $ISO
(1,4) \equiv SO (1,4) \otimes_{s} R^{5}$.}} We want to emphasize
here that although we have a natural five-dimensional Minkowski
geometry to realize the new relativity, the fifth dimension here
should be considered neither as space nor time ($\sigma$ has {the}
dimension of time/mass). In the following section, we will try to
explore the physics meaning of this extra coordinate from two
different points of view.

Before ending this section,  let us consider, for reasons to be
clarified below, one more deformation in relativity by imposing a
third invariant. Here, there is even less physical guidance on what
should be the appropriate quantity to consider, but an infrared
bound seems to be quite meaningful  \cite{tsr,spha}. As we will
argue, it can, for example,  end the many possible iterative
deformations that can be introduced along these lines.  As in the
earlier discussions, let us introduce a parameter $\rho$ external to
the five-dimensional manifold that we have been considering {and}
for which the coordinates are $(x^{0}=c\,t, x^{i}, x^{4}=\kappa
c\,\sigma)$. Following the discussion of Galilean boost, let us
introduce special translations of the form
\be x^{\ssc A}\to x^{\ssc
A} + \Delta x^{\ssc A}\;,\quad \Delta x^{\ssc A} (\rho) = {\mathcal
V}^{\ssc A} \rho \;,        \label{vA}
\ee
where we have identified
${\mathcal V}^{\ssc A}  = \frac{dx^{\ssc A}}{d\rho}$. The
generators, $O_{\!\ssc A}^{\prime}$, of these translations will obey
the same commutation relations as the conventional five-dimensional
translation generators. Furthermore, let us extend the 5D manifold
by including this new coordinate $\rho$ as $x^{\ssc\mathcal  M} =
(x^{\ssc A}, \ell\rho), {\mathcal M} = 0,1,2,3,4,5$ where $\ell$ is
an invariant length. With this choice of the fifth coordinate we see
that the new coordinate $\rho$ is dimensionless. As a result,
${\mathcal V}^{\ssc A}$ defined in Eq.(\ref{vA}) has the dimension
of length and we identify this with a coordinate vector of
translation ${\mathcal V}^{\ssc A}  = z^{\ssc A}$. (The meaning of
this vector will be discussed in Sec.V.) As in Galilean relativity,
we see that $z^{\ssc A}$ can take any arbitrary value. However,
following the earlier discussion on special relativity, let us
introduce a coordinate vector on this six dimensional manifold as
\be
X^{\!\ssc\ssc\mathcal M} = (X^{\!\ssc A}, X^{\ssc 5})
 = \left(\frac{z^{\ssc A}}{\sqrt{\ell^{2} -  \eta_{\ssc A\!B}z^{\ssc A}z^{\ssc B}}},
\frac{\ell}{\sqrt{\ell^{2} - \eta_{\ssc A\!B}z^{\ssc A}z^{\ssc B}}}\right)
 = \left(G\gamma^{\ssc A}, G\right)\;,      \label{X0}
\ee
with $\gamma^{A} = z^{A}/\ell$. It is clear that this coordinate vector will satisfy the condition
\be
\eta_{\ssc\mathcal  M\!\mathcal N} X^{\!\ssc\mathcal M}X^{\!\ssc\mathcal N} = \eta_{\ssc A\!B} X^{\!\ssc A}X^{\!\ssc B}
- \left(X^{\!\ssc 5}\right)^{2} = -1 \;,        \label{X}
\ee
which, in turn, will imply that
\be
\eta_{\ssc A\!B} z^{\ssc A}z^{\ssc B} = \ell^{2} \left(1 - \frac{1}{G^{2}}\right) \leq \ell^{2} \;.
\label{zbound}
\ee
The constraint  introduces an invariant length scale into the
problem. This construction also makes clear that the special
five-dimensional translations $z^{\ssc A}$ live on the coset
$z^{\ssc A} \in ${\small \boldmath $SO(1,5)/SO(1,4)$}, instead of
$z^{\ssc A} \in ${\small\boldmath $R^5$} and the enlarged de Sitter
group {\small \boldmath$SO(1,5)$} corresponds to the symmetry of
this deformed relativity.  The new translations can now be thought
of as a new kind of boost described by the generators
\be \label{tboost} J_{\!\ssc A5}  \equiv O_{\!\ssc A}^{\prime} = i\,
(x_{\!\ssc A}
\partial_{\ssc 5} -x_{\ssc 5}\, \partial_{\!\ssc A}) \;, \ee with
\be [J_{\!\ssc A\!B}, O_{\!\ssc C}^{\prime} ] =
 i \,(\eta_{\ssc B\!C} O_{\!\ssc A}^{\prime}  - \eta_{\ssc A\!C} O_{\!\ssc B}^{\prime} ) \;.
\ee
The deformation relative to the {\small\boldmath $ISO(1,4)$}  algebra is now obtained to be
\be
[O_{\!\ssc A}^{\prime} , O_{\!\ssc B}^{\prime} ]  \longrightarrow  i\, J_{\ssc A\!B} \;,\label{O'}
\ee
for $A,B=0, 1,2,3,4$. Indeed, all the symmetry generators can be written as
\be \label{15}
J_{\ssc\mathcal  M\!\mathcal N}
= i\, (x_{\!\ssc\mathcal  M} \,\partial_{\!\ssc\mathcal N}
 -x_{\!\ssc\mathcal  N}\, \partial_{\!\ssc\mathcal  M}) \;,
\ee
for ${\mathcal M,\mathcal N}= 0,1,2,3,4,5$, giving the linear realization of the {\small \boldmath $SO(1,5)$}
symmetry. This is  what we consider to be the true (full) symmetry of quantum relativity.
Discussions in the rest of the paper will {attempt to} justify this choice as a sensible one.

We note that because of the constraint of Eq.(\ref{X}), the relevant
part of the six-dimensional geometry is actually a five-dimensional
hypersurface given by \be \eta_{\ssc\mathcal  M\!\mathcal N}
X^{\!\ssc\mathcal  M}X^{\!\ssc\mathcal  N} = -1\;.       \label{X1}
\ee (As we discuss later in Sec.V, the coordinates $x^{\ssc\mathcal
M}$ and $X^{\!\ssc\mathcal  M}$ give different  parametrizations of
a point in this six dimensional manifold.) The five-dimensional
hypersurface does not admit simple translational symmetry along any
of the six coordinates anymore.  Hence, the above scheme of
relativity deformations naturally ends here. To be more specific, it
ends once we put in a deformation that imposes an invariant bound on
the displacement vector  generalizing the space-time coordinate
itself. The transformations generated by operators in
Eq.(\ref{tboost}) are isometries of the 5D hypersurface mixing
$x^{\ssc 5}$ with the other coordinates. We call them (de Sitter)
translational boosts. Taking this as the quantum relativity forces
us to consider the 5D hypersurface $\mbox{dS}_{\ssc 5}$, which is a
de Sitter space compatible with a positive cosmological constant, as
the arena for (quantum) space-time. However, we  not only have
$\mbox{dS}_{\ssc 5}$ having one more dimension than the
$\mbox{dS}_{\ssc 4}$ curved space-time conventionally considered in
the cosmology literature, but quantum relativity also suggests that
the extra coordinates $x^{\ssc 4}$ and $x^{\ssc 5}$ are quite
different from the conventional (spacelike) space-time coordinates.

\section{Some Physics of the Momentum Boosts and the $x^{\ssc 4}$ coordinate}
In this section, we focus on the relativity with only one extra
invariant scale, $\kappa c$. Discussions here can be considered as
{relevant} only to the intermediate case without the  last
deformation involving the invariant length $\ell$. The deformed
relativity up to the level of {\small \boldmath $SO(1,4)$} is
essentially the same as the DSR constructions, with a different
{parametrization for} the energy-momentum surface defined by
Eq.(\ref{pi2})  \cite{KNds}. The linear realization of the
transformations presented here has been discussed implicitly, most
notably in Ref.\cite{GL}, although the physics involved is not
clearly discussed .  We view the transformations here as what they
should be, namely, transformations of (quantum) reference frames, in
order to extract a better understanding of a sensible interpretation
of physics  issues involved. We want to emphasize right away that
the deformation, introducing the new momentum boosts as distinct
from the Lorentz boosts in the linear realization through the
5-geometry, is characterized by a central idea contained essentially
in the  defining relation $p^\mu\equiv\frac{dx^\mu}{d\sigma}$. This
is nothing less than introducing {\em  a new definition of the
energy-momentum} 4-vector, whose implications will be discussed
{later} in this section. To emphasize, we note that $\sigma$ (or
$x^{\ssc 4}$) as a coordinate is external to the four-dimensional
space-time, and hence $p^\mu$ so defined is different from the old
definition of (Einstein) energy-momentum in the 4D space-time. {\em
In fact, we consider it necessary to take special caution against
thinking of  $x^{\ssc 4}$ as simply an extra space-time coordinate.
}

The new relativity (in this intermediate case) only adds a new
dimension parametrized by  $x^{\ssc 4}$ and we note that the set of
Lorentz transformations continue to be a part of the isometry group
of the extended 5D manifold characterizing rotations within any 4D
space-time sub-manifold. However, there are also new symmetry
transformations in this 5D manifold. These are the momentum boosts.
To better appreciate the physics of the momentum boosts generated by
the $O_{\!\mu}$'s, let us analyze the finite transformations under
such boosts and, in particular, examine the transformation of the
energy-momentum 4-vector. To keep the discussion parallel to what we
know in Einstein relativity, let us summarize some of the essential
formulae from the latter.We recall that in an inertial frame
characterized by the velocity $\vec{\beta} = \vec{v}/c$ (note that
we use the  $\vec{\cdot}$ notation in this paper to denote a generic
vector defined on a manifold of any dimension; whenever {ambiguities
are likely to arise} , we will use a notation {such as}
$\vec{\cdot}^{\,\,\ssc n}$ to {define} explicitly the
{dimensionality of the} vector), the coordinates transform as
\begin{eqnarray}
x^{\prime\ssc\, 0} & = & \gamma \left(x^{\ssc 0} - \vec{\beta}\cdot \vec{x}\right),  \nonumber\\
\vec{x^{\prime}}  & = & \vec{x} + \vec{\beta}
\left(\frac{(\gamma - 1)}{\beta^{2}} \vec{\beta}\cdot \vec{x} - \gamma x^{\ssc 0}\right)\;, \label{boost}
\end{eqnarray}
where $\beta^{2} = \vec{\beta}\cdot \vec{\beta}$ and $\gamma = 1/\sqrt{1-\beta^{2}}$
denotes the Lorentz contraction factor. Furthermore, if we have a particle moving with a
velocity $\vec{\beta_{\ssc 1}} = \vec{v_{\!\ssc 1}}/c$, then in the boosted frame, it will have
a velocity given by
\begin{equation}
\vec{\beta_{1}^{\,\prime}} = \frac{\gamma^{-1}}{1 - \vec{\beta}\cdot \vec{\beta_{\ssc 1}}} \left[\vec{\beta_{\ssc 1}}
 + \vec{\beta}\left(\frac{(\gamma - 1)}{\beta^{2}} \vec{\beta}\cdot \vec{\beta_{\ssc 1}} - \gamma\right)\right]\;.
\label{velocity}
\end{equation}
This gives the formula for the composition of velocities and, in
particular, when $\vec{\beta} = \beta (1,0,0)$ and $\vec{\beta_{\ssc
1}} = \beta_{\ssc 1} (1,0,0)$, reduces to the well-known formula
\begin{equation}
\beta_{\ssc 1}^{\,\prime} = \frac{\beta_{\ssc 1} - \beta}{1 - \beta\beta_{\ssc 1}}\;,
\end{equation}
which can also be written as
\begin{equation}
v_{\!\ssc 1}^{\prime} = \frac{v_{\!\ssc 1} - v}{1 - \frac{vv_{\!\ssc 1}}{c^{2}}}\;. \label{velocity1}
\end{equation}

The momentum boosts of the five-dimensional  geometry generated by
$O_{\!\mu}$ can also be understood along the same lines.  If we
consider an inertial frame characterized by the momentum 4-vector
$\vec{\alpha} = \vec{p}/\kappa c$, then the 5D coordinates will
transform as
\begin{eqnarray}
x^{\,\prime\, \ssc 4} & = & \Gamma \left(x^{\ssc 4} - \vec{\alpha}\cdot  \vec{x}\right)\;,      \nonumber\\
x^{\,\prime\, \mu} & = & x^{\mu} + \alpha^{\mu}\left(\frac{(\Gamma - 1)}{\alpha^{2}}
\vec{\alpha}\cdot  \vec{x} - \Gamma x^{\ssc 4}\right)\;,        \label{pboost}
\end{eqnarray}
where $\vec{\alpha}\cdot  \vec{x}= \eta_{\mu\nu} \,\alpha^{\mu}x^{\nu}$,
$\alpha^{2} = \vec{\alpha}\cdot  \vec{\alpha}$, and $\Gamma = 1/\sqrt{1 - \alpha^{2}}$
is the analogous ``contraction factor". It can be checked easily that under these
transformations, the metric of the manifold remains invariant, namely,
\begin{equation}
\eta^{\prime\,\ssc A\!B} = \eta^{\ssc A\!B}\;,\qquad \eta_{\ssc A\!B}^{\prime} = \eta_{\ssc A\!B}\;,
\end{equation}
so that the new transformations correspond to isometries of the manifold.

Furthermore, if we have a particle moving with a momentum
$\vec{{\alpha}_{\!\ssc 1}} = \vec{{p}_{\!\ssc 1}}/\kappa c$, then in
the  momentum boosted frame it will have a momentum (see
Eq.(\ref{velocity}))
\begin{equation}
\vec{\alpha^{\prime}_{\!\ssc 1}} = \frac{\Gamma^{-1}}{1 - \vec{\alpha}\cdot \vec{\alpha_{\!\ssc 1}}}
\left[\vec{\alpha_{\!\ssc 1}} + \vec{\alpha}
\left(\frac{(\Gamma - 1)}{\alpha^{2}} \vec{\alpha}\cdot \vec{\alpha_{\!\ssc 1}} - \Gamma\right)\right]\;.       \label{alpha}
\end{equation}
This gives the formula for the composition of momentum under momentum boosts
and can also be written equivalently as
\begin{equation}
\vec{p^{\prime}_{\!\ssc 1}} = \frac{\Gamma^{-1}}{1- \frac{\vec{p}\cdot \vec{p_{\!\ssc 1}}}{\kappa^{2}c^{2}}}
\left[ \vec{p_{\!\ssc 1}} + \vec{p} \left(\frac{(\Gamma - 1)}{p^{2}} \vec{p}\cdot \vec{p_{\!\ssc 1}} - \Gamma\right)\right]\;.      \label{p}
\end{equation}
In particular, if we consider a momentum boost along the $x^{\ssc
0}$ direction generated by $O_{\!\ssc 0}$ characterized by $\vec{p}
= p (1,0,0,0)$, then the composition of the momentum given by
Eq.(\ref{p}) leads to
\begin{eqnarray}
p_{\!\ssc 1}^{\prime\, \ssc 0} & = & \frac{p_{\!\ssc 1}^{\ssc 0} - p}{1 - \frac{pp_{\!\ssc 1}^{\ssc 0}}{\kappa^{2}c^{2}}}\;,\nonumber\\
\vec{p^{\prime}_{\!\ssc 1}}^{\ssc 3} & = &
 \frac{\sqrt{1 - \frac{p^{2}}{\kappa^{2}c^{2}}}}{1 - \frac{pp_{\!\ssc 1}^{\ssc 0}}{\kappa^{2}c^{2}}}\, \vec{p_{\!\ssc 1}}^{\ssc 3}   \;,    \label{0boost}
\end{eqnarray}
which can be compared with the {formula Eq.(\ref{velocity1}) for}
velocity composition in Einstein relativity. Furthermore, if we
assume that the  particle characterized by a rest mass $m_{\ssc 1}$
is in its rest frame so that  $\vec{p_{\!\ssc 1}} = m_{\ssc 1}c
(1,0,0,0)$ and the momentum boost of the form $\vec{p} = m
c(1,0,0,0)$, then Eq.(\ref{0boost}) leads to the composition law
\begin{equation}    \label{m}
m_{\ssc 1}^{\prime} = \frac{m_{\ssc 1} - m}{1 - \frac{mm_{\ssc 1}}{\kappa^{2}}} \;.
\end{equation}
An Einsteinian particle with the rest mass $m_{\ssc 1}$ has momentum
that can be parametrized as $\vec{p_{\!\ssc 1}} = (\gamma m_{\ssc
1}\,c, \gamma m_{\ssc 1}\,c \, \beta_{\ssc 1}^{\,i})$ which
satisfies the on shell condition (a terminology of relativistic
quantum field theory)
\[
\eta_{\mu\!\nu} p_{\!\ssc 1}^{\mu}p_{\!\ssc 1}^{\nu} = m_{\ssc 1}^{2}c^{2}  \;,
\]
in any Lorentzian frame. In particular, there is the particle rest
frame in which we have $\vec{p_{\!\ssc 1}} {=} (m_{\ssc
1}\,c,0,0,0)$. We {normally} think  this {as} the reference frame
defined by the particle itself; or the {the frame where} particle is
the observer. {As a result,} the particle does not see its own
motion, but does see its own mass or energy. The introduction of the
momentum boosts relating {different} reference frames  generalizes
that perspective. Just as $\beta_{\ssc 1}^2$ is not invariant under
Lorentz boosts, {$p_{\!\ssc 1}^2$} is {also} not invariant under the
momentum boosts. {Furthermore, just as} there is the preferred rest
frame {with $\beta_{\ssc 1}^2=0$,} or equivalently {with} the
4-velocity  characterized by {$\vec{u_{\ssc 1}} = (1,0,0,0)$},
{similarly} the linearly realized DSR introduces a preferred
``particle" frame with {$p_{\!\ssc 1}^2=0$}, {which is equivalently}
characterized by the 5-momentum {$\vec{\pi_{\!\ssc 1}}
=(0,0,0,0,1)$} as {obtained from Eq.(\ref{m}) by} setting
{$p_{\!\ssc 1}^{\mu} = p^{\mu}$}.  This is the ``true particle
frame" in which the ``particle" does not see itself, neither its
motion nor its mass/energy. The rest frame of Einstein relativity is
only the frame that has no relative motion {with respect} to the
particle.

As we have mentioned earlier, we would like to view the momentum
boosted frames as quantum reference frames.  We will see here that
the interpretation is in a way necessary, as we look into how other
momentum 4-vectors look like in such a reference frame. Looking at
Eq.(\ref{0boost}) we see that
\begin{equation}
\eta_{\mu\!\nu} \,p_{\!\ssc 1}^{\prime\mu}  p_{\!\ssc 1}^{\prime\nu}
= \frac{1}{\left(1 - \frac{pp_{\!\ssc 1}^{\ssc 0}}{\kappa^{2}c^{2}}\right)^{2}}
 \left[(p_{\!\ssc 1}^{\ssc  0} - p)^{2} - \left(1 - \frac{p^{2}}{\kappa^{2}c^{2}}\right)
   (\vec{p_{\!\ssc 1}}^{\ssc 3})^2 \right]
\neq \eta_{\mu\!\nu}\, p_{\!\ssc 1}^{\mu} p_{\!\ssc 1}^{\nu} \;.        \label{massshell}
\end{equation}
where $(\vec{p_{\!\ssc 1}}^{\ssc 3})^2$ is the magnitude of the
momentum 3-vector. Because of the complicated dependence on $c$ in
here, we cannot even write $\eta_{\mu\nu}  \, p_{\!\ssc 1}^{\prime
\mu}  p_{\!\ssc 1}^{\prime\nu} = m_{\ssc 1}^{\prime\,2} \, c^{2}$.
The quantum field theoretical concept of off-shellness is what we
consider applicable here. Quantum states are either on shell or off
shell, as observed from a classical frame. When boosted to a quantum
frame characterized by even an on shell quantum state, the state
does not observe itself, and observes {the} other originally on
shell states as generally off shell. The concept of {an} off shell
{state} is related to the uncertainty principle. Unlike a classical
particle, a quantum state, even on shell,  {has associated}
uncertainties. If such a state is to be taken as the reference
frame, or observer (measuring apparatus), it is very reasonable to
expect {that}  the apparatus imposes its own uncertainties onto
whatever it {observes/measures}.  We have to be cautious here though
about whether a quantum measuring apparatus has any practical
possibility of being realized. After all, the only true practical
observers, {namely, human beings,} are basically classical.

We see that the conceptually small step that we take here  is indeed
a bold one. Our explicit formulation of the natural linear
realization of the momentum boosts of DSR  looks highly
unconventional. It begs the question  if a consistent and viable
phenomenological interpretation exists --- a question {for which} we
are only {going to provide here} some partial {answer in the}
affirmative. In Einstein relativity, we have $\eta_{\mu\!\nu} p^\mu
p^\nu = m^2 c^2$ and $p^\mu=m c\, u^\mu$. In quantum physics, we are
familiar with the concept of off shell states which violate the
first equation. Our explicit analysis of the momentum boost
illustrates that the on shell condition is not preserved under a
momentum boost. The momentum boost analyzed above is one of the
simplest (namely, a boost along the $x^{\ssc 0}$ axis), but our
conclusions obviously hold for any other more complicated momentum
boost. In fact, it is easily derived from Eq. (\ref{p}) that an
arbitrary momentum boost leads to
\begin{equation}
\eta_{\mu\nu} \, p_{\!\ssc 1}^{\prime\,\mu} p_{\!\ssc 1}^{\prime\,\nu}
= \frac{1}{\left(1 - \frac{\vec{p}\cdot \vec{p}_{\!\ssc 1}}{\kappa^{2}c^{2}}\right)^{2}}
\left[(\vec{p}_{\!\ssc 1} - \vec{p})^{2} + \frac{1}{\kappa^{2}c^{2}}
\left((\vec{p}\cdot \vec{p}_{\!\ssc 1})^{2} - p^{2} p_{\!\ssc 1}^{2}\right)\right]
\neq \eta_{\mu\nu} p_{\!\ssc 1}^{\mu} p_{\!\ssc 1}^{\nu} \;.
\end{equation}
Rather the condition $p^\mu=m c\, u^\mu$ is actually given up  right at the beginning
of the formulation. The linear realization of momentum boosts as distinct from velocity
(Lorentz) boosts introduces the defining relation
$p^{\mu} = \frac{dx^{\mu}}{d\sigma} = \kappa c \frac{dx^{\mu}}{dx^{\ssc 4}}$
{in contrast to the Einstein relativity} limit of $m c\, u^\mu = m  \frac{dx^{\mu}}{dx^{\tau}}$
as $m c \frac{dx^{\mu}}{dx^{\ssc 0}}$.

So far, we have not clarified the nature of the extra coordinate
$\sigma$. This leads to $x^{\ssc 4} = \kappa c \sigma$ which is
necessitated by the desire to have a bound on the energy-momentum
4-vector.  This new coordinate has a spacelike signature, but has
the dimensions of time/mass. This suggests that from the physics
point of view it has the character of time as opposed to space
(which its signature will suggest). In fact, $\sigma$ should be
considered neither as space nor as time in this 5D space. In this
sense, the frames in this 5D manifold that we are analyzing should
be considered different from the ones that arise in a naive 5D
extension of the usual 4D space-time by adding an extra spatial
dimension. To appreciate this a bit more and also to bring out the
nature of the coordinate $\sigma$, let us consider the following.
Let us recall that in deforming Galilean relativity to Einstein
relativity, one introduces Lorentz boosts which mix up space and
time and are characterized by a velocity. The {(instantaneous)}
velocity of a particle is defined as $v^{i} = \frac{dx^{i}}{dt} = c
\frac{dx^{i}}{dx^{\ssc 0}}$. This three-dimensional velocity, of
course, does not transform covariantly under a Lorentz boost.
However, the extra coordinate $x^{\ssc 0}$ is in a way already there
as the time parameter $t$ and the velocity is also there as the time
derivative in the 3D Galilean theory. In a parallel manner, in
deforming Einstein relativity to DSR, we introduce momentum boosts
which mix up the extra coordinate with the 4D coordinates and are
characterized by a momentum. In this case, the momentum of a
particle is then to be defined as $p^{\mu} =
\frac{dx^{\mu}}{d\sigma} = \kappa c \frac{dx^{\mu}}{dx^{\ssc 4}}$.
This momentum does not transform covariantly under momentum boosts
(see Eq.(\ref{p})). However, unlike the case of deforming the
Galilean velocity boosts to Lorentzian ones, here the extra
coordinate $x^{\ssc 4}$ or  $\sigma$ as a parameter is new, and the
definition of the momentum as a $\sigma$ derivative does not
coincide with the {conventional} definition of the momentum in
Newtonian physics or Einstein relativity. In the latter case, one
defines $p_{\mbox{\tiny (ER)}}^{\mu} = m \frac{dx^{\mu}}{d\tau}$
where $\tau$ represents the proper time. We know that  the
{definition of}  $p_{\mbox{\tiny (ER)}}^{\mu}$
 and its Newtonian limit are valid  concepts. So, we do need to reconcile
the two definitions of momentum {in} the regime of Einstein
relativity. That can be achieved if we identify $\sigma =
\frac{\tau}{m}$ which has the dimension of time/mass as we have
pointed out earlier. With this identification we see that the
$\sigma$ coordinate for a (classical) particle observed from a
classical frame is essentially the Einstein proper time.  {(}In
fact, to the extent that {the definition of} $p_{\mbox{\tiny
(ER)}}^{\mu}$  is valid in quantum mechanics, we expect the relation
to be valid to a certain extent even for quantum states observed
from a classical frame.{)} The mass factor, however, gives particles
in the state of motion different $\sigma$ locations. Note that such
an identification holds only for $m\neq 0$ {just as the
identification} $p_{\mbox{\tiny (ER)}}^{\mu} = m
\frac{dx^{\mu}}{d\tau}$ {holds only when} $m\neq 0$. For the
massless photon, for example,  $p_{\mbox{\tiny (ER)}}^{\mu}$
{instead}
 depends on the photon frequency and wavelength -- {an idea with origin in}
quantum physics.That may actually be taken as a hint that the
classical notion of $p_{\mbox{\tiny (ER)}}^{\mu} = m \frac{dx^{\mu}}{d\tau}$
is not fully valid in true quantum physics.

Finally, we comment briefly on the $O_i$ momentum boosts here. Such
a boost is one characterized, for example, by a relative
energy-momentum $\vec{p}= (0,p,0,0)$ which cannot correspond to an
on shell state observed from the original classical reference frame.
However, if momentum boosts can take us to a quantum reference frame
{and} change the observed on shell nature of states, there is no
reason why one cannot accept the reference frame itself to be
characterized by an off shell energy-momentum vector either. Physics
described from such a frame certainly looks peculiar, though it may
not concern a classical {human} observer.

\section{On the Full Quantum Relativity and Phase-Space Symmetry}
There has already been a lot of work and discussion on the subject of DSR and TSR
(doubly and triply special relativity). In this section, we summarize
some of the results on the symmetry aspects that exist in the literature while
 adapting them into our point of view. In particular, we would be
interested in the formulations of TSR in Ref.\cite{tsr,CO} which is
proposed as the full quantum relativity. The results of
Ref.\cite{tsr}  start with the ``phase-space" algebra of DSR with
noncommutative space-time coordinate, which can be written according
to our conventions as
\bea
&& [M_{\mu\nu}, M_{\lambda\rho}] =
i(\eta_{\nu\!\lambda} M_{\mu\rho} - \eta_{\mu\!\lambda} M_{\nu\rho}
+ \eta_{\mu\rho} M_{\nu\lambda} - \eta_{\nu\rho} M_{\mu\lambda})\;,
\nonumber\\
&& [M_{\mu\nu}, \hat{P}_{\!\lambda}] =  i \,(\eta_{\nu\!\lambda} \hat{P}_{\!\mu}
    - \eta_{\mu\!\lambda} \hat{P}_{\!\nu}) \;,
\nonumber \\
&& [M_{\mu\nu}, \hat{X}_{\!\lambda}] =  i \,(\eta_{\nu\!\lambda} \hat{X}_{\!\mu}
    - \eta_{\mu\!\lambda} \hat{X}_{\!\nu}) \;,
\nonumber \\
&& [\hat{X}_{\mu}, \hat{X}_{\!\nu}] =  \frac{i}{\kappa^2 c^2}  M_{\mu\nu} \;,
\nonumber \\
&& [\hat{P}_{\mu}, \hat{P}_{\!\nu}] =  0 \;,
\nonumber \\
&& [\hat{X}_{\mu}, \hat{P}_{\!\nu}] = - i\, \eta_{\mu\nu} \;.
 \label{tsr1}
\eea
This is identified as the algebra of DSR phase-space symmetry
and we note that  $\hat{X}$ and $\hat{P}$ correspond to generators
(operators) of the algebra in contrast to the $x$ and $p$
coordinates described in the earlier sections. A second deformation
of Eq.(\ref{tsr1}) is considered with a view to implement the third
invariant as a length $\ell$ related to the cosmological constant
($\Lambda=\ell^{-2}$). In this case the commutator of {the} momentum
operators in Eq.(\ref{tsr1}) is deformed to \be [\hat{P}_{\mu},
\hat{P}_{\!\nu}] =   \frac{i}{\ell^2}  M_{\mu\nu} \;. \ee However,
this deformation leads to a violation {of} the Jacobi identity which
{induces} a further  modification of  the Heisenberg commutator
(between $\hat{X}$ and $\hat{P}$) involving  a complicated
(quadratic) expression in the generators. This algebra is  then
identified as the quantum algebra (of TSR).  We also note here that
it was pointed out in Ref.\cite{tsr} that this  algebra can be
represented in terms of  coordinates and derivatives of a six-dimensional
manifold. It is important to recognize that in both
cases (DSR and TSR), in addition to the deformations, the usual 4D
coordinates are promoted to generators of the algebra, also to be
interpreted as representing a noncommutative geometry of quantum
space-time.

At this point, it is interesting to compare the original DSR to TSR
deformation of Ref.\cite{tsr} with our formulation of the quantum
relativity algebra. After the deformation from Einstein relativity
to {\small \boldmath $SO(1,4)$}, we have a linear realization of the
algebra at the level of DSR. Our relativity algebra is set on a 5D
commutative manifold. We do not have coordinate operators as
generators of the algebra. However, it is worth noting that the four
generators, $O_{\!\mu}$'s, generating momentum boosts satisfy the
same commutation relations as the $\hat{X}_{\mu}$ operators in
Eq.(\ref{tsr1}) (see, for example, Eqs.(\ref{O}) and (\ref{O1})).
Therefore, formally we can identify $O_{\!\mu}$ as $-{\kappa
\,c}\,\hat{X}_{\mu}$ with the explicit form following from
Eq.(\ref{O}) \be \label{Xop} \hat{X}_{\mu} = -\frac{1}{\kappa \,c}\,
i\, (x_\mu \partial_{\ssc 4} -x_{\ssc 4}\, \partial_{\mu}) \;, \ee
which actually seems like a very reasonable ``quantum"
generalization of the classical, or rather Einstein, space-time
position. In the limit ${\kappa \,c}\,\to \infty$ (with $\sigma\to
0$ as $1/\kappa c$) and $ i\,\partial_{\ssc 4}\equiv p_{\ssc
4}=-{\kappa \,c}$ ($p_{\ssc 4}=\eta_{\ssc 4A}p^{\ssc A}$), the
operators reduce to $x_\mu$. Such an identification as in
Eq.(\ref{Xop}), provides a bridge between a noncommutative geometric
description of 4D quantum space-time (details of which
 await {further} investigation) and {our perspective of} quantum relativity  as linearly realized
on a 5D, or eventually 6D, commutative manifold --- a geometric description
beyond the space-time perspective. We believe the two pictures to be complementary.

The symmetry can now be enlarged  to {\small \boldmath $ISO(1,4)$}
by incorporating translations of the 5D manifold.  The five of them
are $p_{\!\ssc A}\equiv i\,\partial_{\!\ssc A}$. Dropping $p_{\ssc
4}$ for the moment, they almost satisfy the full ``phase-space"
algebra of DSR. The only problem comes from the Heisenberg
commutator which is no longer canonical. While one can reasonably
argue that our linear realization of the DSR simply suggests that
the Heisenberg commutator should be modified, it is not what we want
to focus on here.  We take the full quantum relativity at the next
(TSR) level, namely, with the third invariant $\ell$. Again,
forgetting the $\hat{X}_\mu$-$\hat{P}_{\mu}$ commutator for the
moment, the algebra which is essentially {\small \boldmath
$ISO(1,4)$} is deformed to {\small \boldmath $SO(1,5)$}. In this
case, it is possible to formally identify ${\ell}\,\hat{P}_{\mu}=
O_{\!\mu}^{\prime} \;(=J_{\mu\ssc 5})$ (see Eqs.(\ref{tboost}) and
(\ref{O'})) which gives the  explicit linear realization \be
\label{Pop} \hat{P}_{\mu} = \frac{1}{\ell}\, i\, (x_\mu
\partial_{\ssc 5} -x_{\ssc 5}\, \partial_{\mu}) \;.
\ee
Once  again
this gives a very reasonable ``quantum"  generalization of the
``classical" $p_\mu$. This is seen by noting that in the limit $\ell
\,\to \infty$ (with $\rho\to 0$ as $1/\ell$) , $\hat{P}_{\mu}$
reduces to $ i\,\partial_{\mu}\equiv p_{\mu}$. All of these fit in
very nicely, except for the $\hat{X}_\mu$-$\hat{P}_{\mu}$ commutator
and the issue of the extra $\hat{P}_{\ssc 4}$, or rather $O_{\!\ssc
4}^{\prime}$ which has not been addressed so far.

The missing link in the above discussion actually can be obtained
from the analysis in Ref.\cite{CO} which restores the Lie-algebraic
description of the TSR algebra by identifying the right-hand side of
the  Heisenberg commutator as a central charge generator $\hat{F}$
of the original algebra with {relevant} commutators
 also deformed, yielding
\be \label{Fop}
  [\hat{X}_{\mu}, \hat{P}_{\!\nu}] = -i \, \eta_{\mu\nu} \hat{F} \;,
\qquad
 [\hat{X}_{\mu}, \hat{F}] = \frac{i}{\kappa^2 c^2} \hat{P}_{\mu} \;,
\qquad [\hat{P}_{\mu}, \hat{F}] = -\frac{i}{\ell^2} \hat{X}_{\mu}
\;.
\ee
This identifies the resulting algebra exactly with {\small
\boldmath $SO(1,5)$}, and, {therefore}, $\hat{F}$ loses its
character as a central charge generator. It is also interesting that
this algebra has been identified as the mathematical stabilizer of
the ``Poincar\'e+Heisenberg" symmetry  \cite{CO}. The generator
$\hat{F}$ is essentially $O_{\!\ssc 4}^{\prime}$ (or equivalently
$\hat{P}_{\ssc 4}$). Explicitly, $O_{\!\ssc 4}^{\prime}= J_{\!\ssc
45} = -\kappa c \hat{F}$.

Nonlinear realizations of the different versions of quantum
relativity focus on the description of the (four) space-time
geometry as a noncommutative geometry. Here, the nonvanishing
commutator among $\hat{X}_{\mu}$'s may be considered to result from
the curvature of the energy-momentum space  [{\it cf.}
Eq.(\ref{pi2})] while the  nonvanishing commutators among
$\hat{P}_{\mu}$'s is considered a result of the ``space-time"
curvature \cite{dsr2,M} as characterized by the nonvanishing
cosmological constant. Our formulation through the linear
realization yields an explicit identification of the quantum
operators as generalizations of the familiar phase-space coordinates
variables. This is based on a 6D geometry which is commutative. Both
the extended $x$-space and the $p$-space are copies of $\mbox{dS}_5$
and the isometry of each contains the full phase-space symmetry
algebra of the quantum theory of 4D noncommutative space-time. The
$x^\mu$-variables and the $p^\mu$-variables are on the same
symmetric footing. In our opinion, this is a very {attractive} and
desirable feature. It also clearly suggests that the $x^{\ssc 4}$
and $x^{\ssc 5}$ coordinates should not be interpreted as space-time
coordinates in the usual sense. The generators of the momentum and
translational de Sitter boosts can be viewed as the set of quantum
position and momentum operators.

It is interesting to note that the {\small \boldmath $SO(1,4)$}
algebra, with the $J_{\mu\ssc 4}$ generators taken essentially as
the ``position" variables is identified as the universal basis of a
noncommutative space-time description of various DSR theories
\cite{KNncg}. The full phase-space symmetry algebras, however, {are}
taken to be $\kappa$-Poincar\'e quantum algebras in different bases
\cite{KNncg}. The latter contains $\hat{P}_{\mu}$ generators
extending the {\small \boldmath $SO(1,4)$} algebra with different
deformed commutators from the trivial case of {\small \boldmath
$ISO(1,4)$} corresponding to taking different 4-coordinates for the
coset space  {\small \boldmath $SO(1,4)/SO(1,3)$} for the
energy-momentum  \cite{KNds}. By sticking to the 5-momentum,
$\pi^{\!\ssc A}$, together with a 5D geometric extension of the
space-time, our linear realization allows us to put in the next
deformation of {\small \boldmath $ISO(1,4)$} to the full quantum
relativity of  {\small \boldmath $SO(1,5)$} naturally. The latter
{group} was proposed as the TSR algebra  \cite{tsr} only after the
Lie-algebraic interpretation of Ref.\cite{CO}, by pulling out the
algebraic structure basically {from a deformation of} the
phase-space symmetry algebra directly, as discussed above. With our
formulation of the full quantum relativity of {\small \boldmath
$SO(1,5)$}, it is suggested that not only is the 4D space-time
noncommutativity universal, so is that for the 4D energy-momentum
space. But now both the 6-vectors $x^{\ssc\mathcal M}$ and
$\pi^{\ssc\mathcal M}$ should be living on a $\mbox{dS}_5$. It would
be interesting to see how different choices of 5-coordinates on the
two $\mbox{dS}_5$ or canonical coordinates of the ten-dimensional
``phase-space'' match onto the various DSR theories or other similar
structures for the different nonlinearly realized space-time
structures as well as the role of the related quantum algebras.

\section{The Translational Boosts and De Sitter Geometry}
Similar to the Lorentz and the de Sitter momentum boosts discussed earlier, the
introduction of the de Sitter translational boosts is characterized by the 5-vector
$\vec{z}=\frac{d\vec{x}}{d\rho}=\ell \frac{d\vec{x}}{dx^{\ssc 5}}$. We note
that since the new coordinate $\rho$ is dimensionless, the parameter of boost in this
case carries the same dimension as the coordinates themselves. This should be contrasted
with velocity which represents the parameter of transformation in the case of Lorentz
boosts, and the momentum for the momentum boosts. A generic
$O^\prime$ boost characterized by a vector $\vec{z_{t}}$ can be denoted by the
transformation of the coordinates
\bea
&& x^{\prime\,{\ssc 5}} = G_{\!t} (x^{\ssc 5} - \vec{z_t}\cdot  \vec{x})
= G_{\!t} (x^{\ssc 5} -  \vec{\gamma_t}\cdot \vec{x}) \;,       \nonumber\\
&& \vec{x^{\prime}} = \vec{x} + \vec{\gamma_t}
    \left(\frac{G_{\!t} -1}{\gamma_{t}^2}  \vec{\gamma_t}\cdot \vec{x}  -  G_{\!t} \, x^{\ssc 5} \right)\;,
 \label{dstb0v}
\eea
where we have identified (as defined earlier)
\begin{equation}
\vec{\gamma_t} = \frac{\vec{z_t}}{\ell}\;,
\quad G_{\!t} = \frac{\ell}{\sqrt{\ell^{2} - z_{t}^{2}}}\;,
\quad \vec{\gamma}_{t}\cdot \vec{x} = \eta_{\ssc  A\!B} \gamma_{t}^{\ssc A} x^{\ssc B}\;,
\quad z_{t}^{2} = \ell^{2}\gamma_{t}^{2} = \eta_{\ssc A\!B} z_{t}^{\ssc A} z_{t}^{\ssc B}\;.
\end{equation}
The transformations are analogous to Eqs.(\ref{boost}) and
(\ref{pboost}) representing Lorentz boosts and de Sitter momentum
boosts and the 6D metric tensors $\eta^{\ssc\mathcal M \mathcal N},
\eta_{\ssc\mathcal M \mathcal N }$ are preserved under these
translational boosts.

Any 6-vector and, in particular,  ${X}^{\!\ssc\mathcal  M}$ defined
in Eq.(\ref{X0}) will transform covariantly under a translation
boost as in Eq. (\ref{dstb0v}). However, the 5D vector $\vec{z}$
will transform like the velocity in Eq.(\ref{velocity}) or like the
momentum in Eq.(\ref{alpha}) as
\begin{equation} \label{ztrans}
\vec{\gamma^{\prime}} = \frac{G_{\!t}^{-1}}{1 - \vec{\gamma}\cdot \vec{\gamma_t}}
\left[ \vec{\gamma} + \vec{\gamma_t}
\left(\frac{(G_{\!t} -1)}{\gamma_{t}^{2}} \vec{\gamma}\cdot \vec{\gamma_t} - G_{\!t}\right)\right] \;.
\end{equation}
Further support for identifying the parameter of translational boost
$z^{\ssc A}$ as a coordinate vector can be obtained as follows. Let us note that a point
$x^{\ssc\mathcal  M}$ on the six dimensional manifold satisfying
\begin{equation} \label{ds5}
\eta_{\ssc\mathcal  M\!\mathcal  N} x^{\ssc\mathcal  M} x^{\ssc\mathcal  N} = \eta_{\ssc A\!B} x^{\ssc A} x^{\ssc B} - (x^{\ssc 5})^{2}
 = - \ell^{2} \;,
\end{equation}
can be parametrized alternatively as (for $x^{\ssc 5} > 0$)
\begin{eqnarray}
x^{\ssc A} & = & \ell \omega^{\ssc A} \sinh \zeta\;,    \nonumber\\
x^{\ssc 5} & = & \ell \cosh \zeta\;,    \label{alternate}
\end{eqnarray}
where $\omega^{\ssc A}$ denotes a unit vector on the 5D manifold
satisfying $\eta_{\ssc A\!B}\omega^{\ssc A}\omega^{\ssc B} = 1$. It
is clear now that we can identify the components of
$X^{\!\ssc\mathcal  M}$ in Eq.(\ref{X0}) as
\begin{equation}
G = \cosh \zeta  \;,
\quad z^{\ssc A} = \ell \omega^{\ssc A} \tanh \zeta \;,
\quad \gamma^{\ssc A} = \frac{z^{\ssc A}}{\ell} = \omega^{\ssc A} \tanh \zeta \;,
\quad G^{2} (1 - \gamma^{2}) = 1 \;.        \label{alternate1}
\end{equation}
This brings out the character of  the (alternative) coordinate vector
$z^{\ssc A}$ as a parameter of boost. (This can be contrasted with the angular
representation of a Lorentz boost which has the form
$\gamma = \cosh \theta, |\vec{\beta}| = \tanh \theta, \gamma^{2} (1 - |\vec{\beta}|^{2}) =1$.
 Note the symbol $\gamma$ is used in both cases to represent different
things.) Furthermore, with this identification, we recognize that the 6D vectors $x^{\ssc\mathcal  M}$
and $X^{\!\ssc\mathcal  M}$ can be related simply as
\begin{equation}
x^{\ssc\mathcal  M} = \ell X^{\!\ssc\mathcal  M}.
\end{equation}
In fact, with the identifications in Eqs.(\ref{alternate}) and
(\ref{alternate1}), we note that if we identify
\begin{equation}
\omega^{\ssc A} = \frac{x^{\ssc A}}{\sqrt{x^{2}}},
\end{equation}
where $x^{2} = \eta_{\ssc A\!B}x^{\ssc A}x^{\ssc B}$, we can write
\begin{equation}
z^{\ssc A} = \ell\ \frac{x^{\ssc A}}{\sqrt{x^{2}}}\ \tanh \zeta
= \ell\ \frac{x^{\ssc A}}{\ell \sinh \zeta}\ \tanh \zeta = \frac{x^{\ssc A}}{\cosh \zeta}
= \ell\ \frac{x^{\ssc A}}{x^{\ssc 5}}  \;,
\end{equation}
which is, of course, the definition of Beltrami coordinates for the
de Sitter manifold  dS$_{5}$ (on the Beltrami patch $x^{5}>0$). As
we have  mentioned earlier in connection with Eq.(\ref{X1}), the
basic manifold of our theory is a 5D hypersurface  dS$_{5}$ of the
6D manifold and can be parametrized by five independent coordinates.
The Beltrami coordinates (also known as gnomonic coordinates)
provide a useful coordinate system which preserve geodesics as
straight lines and, therefore, we can use $z^{\ssc A}, A=0,1,2,3,4$
to parametrize our manifold. We note here that there have been some
studies on  de Sitter special relativity  \cite{dssr} (which is
Einstein special relativity formulated on a de Sitter, rather than a
Minkowski space-time) where  Beltrami coordinates are used. Some of
the results obtained there may be used to shed more light on the
physics of quantum relativity.  However, we want to emphasize here
that the (special) quantum relativity that we are discussing here is
{\em not} just a version of de Sitter special relativity. In
particular, as we have discussed, the momentum boost transformations
are expected to relate quantum frames of reference, and one should
be cautious in borrowing physics results from Ref.\cite{dssr}. In
fact, the symmetric role of the relations of $x^{\ssc 4}$ and
$x^{\ssc 5}$ coordinates to the 4D quantum noncommutative position
and momentum operators, respectively, gives a new perspective on
classical de Sitter physics.

Note that the translational boosts are reference frame
transformations that correspond to taking the coordinate origin to a
different location on the dS$_{5}$. The coordinate origin is always
an important part of the frame of reference. The origin is where an
observer measures locations of physical events from. Up to Einstein
relativity, translation of coordinate origin is quite trivial. It
does not change most of the physical quantities like velocity and
energy-momentum measured. However, the situation is a bit different
in the de Sitter geometry. Here, the coordinate origin can be
unambiguously represented by ${z}^{\ssc A}=0$, or location where the
observer measures quantities including location $z^{\ssc A}$ of
events from. The reference frame does not see itself, and must
conclude that its own location is right at the origin of the
coordinate system it measures event locations with. In terms of the
6-coordinate $X^{\!\ssc\mathcal  M}$, the origin has a single
nonvanishing coordinate $X^{\ssc 5}=1$. Consider an event seen as at
location given by a nonvanishing ${z}^{\ssc A}_t\ne 0$, say with
zero velocity and momentum for simplification, transforming to the
new reference frame characterized by the event means translating the
coordinate origin to that location, {\em i.e.} translating by
${z}^{\ssc A}_t$. One can check explicitly that the new coordinates
of the event, $X^{\prime\ssc\mathcal M}$ or $z^{\prime\ssc A}$, as
seen from itself to be obtained from Eqs.(\ref{dstb0v}) and
(\ref{ztrans}) are indeed that of an origin. It also follows that
composing translations gives nontrivial relativistic results beyond
simple addition. In fact, comparing with the velocity and momentum
composition formula of the Lorentz and de Sitter momentum boosts,
respectively ({\em cf.} Eqs.(\ref{velocity},\ref{alpha},\ref{p})),
we have the location composition formula
\begin{equation}
\vec{z^{\prime}_{\ssc 1}} = \frac{G^{-1}}{1- \frac{\vec{z}\cdot \vec{z_{\ssc 1}}}{\ell^{2}}}
\left[ \vec{z_{\ssc 1}} + \vec{z} \left(\frac{(G - 1)}{z^{2}} \vec{z}\cdot \vec{z_{\ssc 1}} - G\right)\right]\;.
\end{equation}
The formula gives the new location 5-vector $\vec{z^{\prime}_{\ssc
1}}$ of an original $\vec{z_{\ssc 1}}$ boosted to a new frame
characterized by  $\vec{z}$ $\left( G^{-2} = 1 - \frac{\eta_{\!\ssc
A\!B} z^{\!\ssc A}z^{\!\ssc B}}{\ell^{2}}\right)$. In particular,
$\vec{z^{\prime}_{\ssc 1}}$ vanishes for $\vec{z}=\vec{z_{\ssc 1}}$.
On the other hand, simple addition of the 6-vector
${X}^{\!\ssc\mathcal  M}$'s does not preserve the de Sitter
constraint of Eq.(\ref{ds5}) characterizing the dS$_5$ hypersurface.
Likewise, conventional 6D translations are not admissible
symmetries.

The dS$_5$ hypersurface is obtained from a 6D manifold with a flat metric,
$\eta_{\ssc\mathcal  M\!\mathcal  N}$, when described in terms of the six coordinates
$x^{\ssc\mathcal  M}$'s. When described in terms of the Beltrami coordinates $z^{\ssc A}$'s,
however, the 5D metric is nontrivial and has the form
\be \label{g}
g_{\ssc \!A\!B}= G^2 \eta_{\ssc A\!B} +\frac{G^4}{{\ell}^2} \eta_{\ssc A\!C}
  \eta_{\ssc B\!D} z^{\ssc C} z^{\ssc D} \;.
\ee
The generators of the isometry group {\small \boldmath $SO(1,5)$}  can be expressed in
terms of these variables as follows. We first note that
$z_{\ssc A}=g_{\ssc \!A\!B}z^{\ssc B}=G^4 \, \eta_{\ssc A\!B} z^{\ssc B},$ (where we have
used the definition $G^{-2} = 1 - \frac{\eta_{\!\ssc A\!B} z^{\!\ssc A}z^{\!\ssc B}}{\ell^{2}}$)
leading to
\be
x_{\!\ssc A}=\frac{1}{G^3}z_{\ssc A}\;, \qquad \mbox{and}\qquad
x_{\ssc 5} = -G\,\ell\;.
\ee
Denoting $i\frac{\partial}{\partial z^{\ssc A}}$ by $q_{\ssc A}$, we have
\bea
p_{\!\ssc A} &=& i\frac{\partial}{\partial x^{\!\ssc A}}
= \frac{i}{G} \frac{\partial}{\partial z^{\ssc A}}
=\frac{1}{G}q_{\ssc A} \;,
\nonumber \\
p_{\ssc 5} &=& i \frac{\partial}{\partial x^{\ssc 5}}
 = \frac{\partial z^{\ssc A}}{\partial x^{\ssc 5}} \ i\frac{\partial}{\partial z^{\ssc A}}
= \frac{\partial z^{\ssc A}}{\partial x^{\ssc 5} }\ q_{\ssc A}
=-\frac{1}{G\,\ell} \, q_{\ssc A} z^{\ssc A}\;.
\eea
Introducing a Lorentzian 5-coordinate $Z_{\!\ssc  A}^{\ssc (\mathcal L)}= G^{-4}\,  z_{\ssc A}
=\eta_{\ssc A\!B} z^{\ssc B}$, we have
\be
[ Z_{\!\ssc  A}^{\ssc (\mathcal L)}, q_{\ssc B} ] =-i \, \eta_{\ssc A\!B} \;.
\ee
A form of Lorentzian `5-momentum' as generators is given by  \cite{Gur}
\be \label{P}
P_{\!\ssc  A}^{\ssc (\mathcal L)} = \frac{1}{\ell} J_{\!\ssc A5}  \;.
\ee
This is in agreement with the noncommutative momentum operator discussed
in Sec.IV above. We have here
\be
P_{\!\ssc  A}^{\ssc (\mathcal L)} = \frac{1}{\ell}
 \left(x_{\!\ssc A}p_{\ssc 5}-x_{\ssc 5}p_{\!\ssc A}\right)
=q_{\ssc A} - Z_{\!\ssc  A}^{\ssc (\mathcal L)}
\; \frac{1}{\ell^2} \left(\eta^{\ssc B\!C} Z_{\!\ssc  B}^{\ssc (\mathcal L)}q_{\ssc C} \right)\;.
\ee
The other ten generators are then given as
\be
J_{\!\ssc A\!B}= Z_{\!\ssc A}^{\ssc (\mathcal L)} q_{\ssc B}
 -Z_{\!\ssc B}^{\ssc (\mathcal L)} q_{\ssc A}
= Z_{\!\ssc  A}^{\ssc (\mathcal L)} P_{\!\ssc B}^{\ssc (\mathcal L)}
 -Z_{\!\ssc  B}^{\ssc (\mathcal L)} P_{\!\ssc  A}^{\ssc (\mathcal L)} \;.
\ee
In fact, one can also write $J_{\!\ssc A5}$ formally as
$Z_{\!\ssc  A}^{\ssc (\mathcal L)} P_{\!\ssc   5}^{\ssc (\mathcal L)}
-Z_{\!\ssc  5}^{\ssc (\mathcal L)} \, P_{\!\ssc  A}^{\ssc (\mathcal L)}$ by taking
$P_{\!\ssc   5}^{\ssc (\mathcal L)}$ as zero (since $Z_{\ssc  5}^{\ssc (\mathcal L)} = -\ell$).
If one further writes  an analogous relation
$P_{\!\ssc 5}^{\ssc (\mathcal L)}= q_{\ssc 5}- \frac{1}{\ell^2} Z_{\!\ssc 5}^{\ssc (\mathcal L)}
 \left(\eta^{\ssc BC}Z_{\!\ssc B}^{\ssc (\mathcal L)}q_{\ssc C} \right)$, it would require that
$q_{\ssc 5}=-\frac{1}{\ell} \left(\eta^{\ssc BC}Z_{\!\ssc B}^{\ssc (\mathcal L)}q_{\ssc C} \right)$.
The latter is, of course, equivalent to $p_{\ssc 5}= \frac{1}{G}\, q_{\ssc 5}$. It is interesting to
note that $\left(\eta^{\ssc BC}Z_{\!\ssc B}^{\ssc (\mathcal L)}q_{\ssc C} \right)$
resembles the conformal symmetry (scale transformation) generator for the 5-geometry
with a Minkowski metric. A further interesting point is that the fifth component of the
`5-momentum' now gives the operator responsible for the new quantum Heisenberg
uncertainty relation as
\[
[\hat{X}_\mu,  \hat{P}_{\nu}] = i \, \eta_{\mu\nu} \frac{1}{\kappa\,c\,\ell}
J_{\!\ssc 45} = -  i \, \eta_{\mu\nu} \left(-\frac{1}{\kappa\,c}
\, P_{\!\ssc  4}^{\ssc (\mathcal L)}\right) \;,
\]
where $\hat{P}_{\nu}=P_\nu^{\ssc (\mathcal L)}$. This follows from
Eq.(\ref{P}) and the result of Sec.IV,  {namely,  from}
Eq.(\ref{Fop}) with $O_{\!\ssc 4}^{\prime}= J_{\!\ssc 45} = -\kappa
c \hat{F}$. The standard Heisenberg expression would be obtained for
$P_{\!\ssc  4}^{\ssc (\mathcal L)}$ taking the value
${-}{\kappa\,c}$. The latter is likely to be admissible as an
eigenvalue condition for the operator; and it invites comparison
with the energy-momentum constraint $p_{\ssc  4}={-}{\kappa\,c}$
before the introduction of the deformation with the translational
boosts. In case $P_{\!\ssc  4}^{\ssc (\mathcal L)}$ takes other
eigenvalues, it could be a generalization of the original
energy-momentum constraint. Hence, the spectrum of $P_{\!\ssc
4}^{\ssc (\mathcal L)}$ as an operator is of central importance.

\section{More on the Momentum in Quantum Relativity}
The interesting thing with the Beltrami coordinate formulation is
that the first five coordinates of the 6-geometry $x^{\ssc A}$ and
hence the coordinates $z^{\ssc A} \;(=x^{\ssc A} \frac{\ell}{x^{\ssc
5}})$ transform as components of coordinate vectors on a 5D space
with Minkowski geometry. Similarly, $Z_{\!\ssc A}^{\ssc (\mathcal
L)}$, $q_{\ssc A}$, and $P_{\ssc A}^{\ssc (\mathcal L)}$, can all be
considered Lorentzian 5-vectors. With the 5-coordinate description
of the $\mbox{dS}_5$ geometry, the  Lorentzian 5-coordinate $Z_{\ssc
A}^{\ssc (\mathcal L)}$
 and the 5-momentum $P_{\ssc A}^{\ssc (\mathcal L)}$ provide
a very nice representation of the relativity symmetry algebra.
Recall that we get to the relativity formulation through
deformations in two steps. The first deformation is introduced by
imposing the Planck scale as a constraint onto the Einstein
4-momentum (Eq.(\ref{Gamma})). This promotes the 4-momentum into a
5-vector $\pi^{\ssc  A}=\Gamma \frac{p^{\ssc  A}}{\kappa\,c}$  with
$p^{\ssc 4}=\kappa\,c$. In terms of the original 6-coordinate
$x^{\ssc\mathcal M}$ and the corresponding
 $p_{\!\ssc\mathcal M}=i\,\partial_{\!\ssc\mathcal M}$,  we also have the nice representation
of the algebra with identification of the noncommutative quantum
space-time position operators within the generators of the algebra
given by Eqs.(\ref{Xop}) and  (\ref{Pop}). There, we have Lorentzian
4-coordinate and 4-momentum $\hat{X}^\mu$ and $\hat{P}_\mu$ with a
quite symmetric role, and an extra $J_{\ssc 45}$ characterizing the
quantum uncertainty relation. The rest of the generators in the
algebra are just the 4D Lorentz symmetry generators. The structure
seems to be depicting a 4D quantum space-time. It is also easy to
see that $\hat{P}_{\!\ssc 4}\,={P}_{\!\ssc 4}^{\ssc (\mathcal L)}$
together with the four $\hat{P}_\mu\, =P_{\mu}^{\ssc (\mathcal
L)}$'s transform as the  5-momentum  of a 5D Lorentzian symmetry,
with now  $\hat{X}^\mu$ forming part of the 5D angular momentum. The
translational boosts mix the new, {sixth momentum component}
$p^{\ssc 5}$ with the other five making a consistent interpretation
of the 5-momentum constraint of Eq.(\ref{pi}) questionable. It is
then very reasonable to expect the momentum constraint to be taken
as one on $P_{\ssc A}^{\ssc (\mathcal L)}$, which even has a natural
extension to have a vanishing sixth component. Hence, we would like
to think about $P^{\ssc (\mathcal L)^{\ssc A}}=\eta^{\ssc A\!B}
P_{\ssc B}^{\ssc (\mathcal L)}$. However, under the usual framework
of dynamics, taking the operator form of the momentum in the
(coordinate representation) to the momentum variable as (particle)
coordinate derivative requires the equation of motion. Let us take a
look at this issue from a different perspective.

We discuss below momentum as coordinate derivative and conserved
vector corresponds to the relativity symmetry. We start from writing
the ``classical angular momenta" in terms of components of the
relevant Beltrami 5-vectors. All the generators of {\small \boldmath
$SO(1,5)$} are expected to correspond to conserved quantities in the
classical sense. We again start with $L^{\ssc\mathcal M \!\mathcal
N}=x^{\ssc\mathcal M} p^{\ssc\mathcal N} - x^{\ssc\mathcal N}
p^{\ssc\mathcal M}$ where the ``classical" momentum $p^{\ssc\mathcal
M}$ is written as $p^{\ssc\mathcal M}=m_{\ssc
\!\Lambda}\frac{dx^{\ssc\mathcal M}}{ds}$. Here,  $m_{\ssc
\!\Lambda}^2$ is a mass-squarelike parameter {\em not} to be taken
directly as the Einstein rest mass-square. It is likely to be some
generalization of the latter. In fact, an apparent natural choice of
the  parameter $m_{\ssc \!\Lambda}^2$ is available from the
eigenvalue of the first Casimir operator for the {$SO(1,5)$}
symmetry \cite{Gur,dssr}.  And the momentum as a coordinate
derivative has, of course, to be defined with respect to the
invariant line element $ds$. In terms of the Beltrami 5-coordinates,
we have \be L^{\!\ssc A5} = - m_{\ssc \!\Lambda} G^2 \ell
\frac{dz^{\ssc A}}{ds} \;. \ee The expression should correspond to
the conserved quantity for the five generators $J_{\!\ssc A5} =\ell
\hat{P}_{\!\ssc A} \; [\mbox{or~} \ell P_{\ssc A}^{\ssc (\mathcal
L)}]$. Hence, we expect the conserved momentum to be given be \be
P^{\ssc A}_{\!\ssc (\mathcal L)}=  m_{\ssc \!\Lambda} G^2
\frac{dz^{\ssc A}}{ds} \;, \ee {\it i.e.,} $L^{\!\ssc A5} = -\ell \,
P^{\ssc A}_{\!\ssc (\mathcal L)}$. It is interesting to note that
$\frac{dP^{\!\ssc A}_{\!\ssc (\mathcal L)}}{ds} =0$ actually is
equivalent to the geodesic equation within $\mbox{dS}_5$
\cite{dssr}. The apparent ``natural" momentum candidate  $q^{\ssc
A}=m_{\ssc \!\Lambda} \frac{dz^{\ssc A}}{ds}$ is related to $P^{\ssc
A}_{\!\ssc (\mathcal L)}$ in a way similar to the relation between
the operator forms of $P_{\!\ssc A}^{\ssc (\mathcal L)}$ and
$q_{\ssc A}$. The coordinate transformation gives $q^{\ssc A}=
\frac{1}{G} p^{\ssc A}$, and hence \be P^{\ssc A}_{\!\ssc (\mathcal
L)}= G^2 q^{\ssc A} -  m_{\ssc \!\Lambda} z^{\ssc A} G \frac{dG}{ds}
\;. \ee Also we have \be L^{\!\ssc A\!B}=z^{\ssc A} G^2 q^{\ssc B}
-z^{\ssc B} G^2 q^{\ssc A} = z^{\ssc A}  P^{\ssc B}_{\!\ssc
(\mathcal L)} -z^{\ssc B} P^{\ssc A}_{\!\ssc (\mathcal L)}\;, \ee
with the same form applicable to $L^{\!\ssc A5}$ by taking $P^{\ssc
5}_{\!\ssc (\mathcal L)}=0$ or equivalently $q^{\ssc 5} = m_{\ssc
\!\Lambda}  \frac{\ell}{G} \frac{dG}{ds}= \frac{1}{G} p^{\ssc 5}$.
The latter result may be written in a form similar to the operator
$q_{\ssc 5}$  given in the previous section, namely  $q^{\ssc 5} =
\frac{G^2}{\ell} \eta_{\ssc A\!B} z^{\ssc A} q^{\ssc B}$. Note that
$g_{\ssc \!A\!B} P^{\ssc A}_{\!\ssc (\mathcal L)} P^{\ssc B}_{\!\ssc
(\mathcal L)}= m_{\ssc \!\Lambda}^2 G^4$.

The above definition of  ``classical" momentum $p^{\ssc\mathcal M}$
of course goes in line with the Newtonian/Einstein starting point of
mass times velocity. However, as discussed in some details in
Sec.III, we have to introduce the new nonquantum-relativistic
Einstein momentum defined as $\frac{dx^\mu}{d\sigma}= \kappa\,c
\frac{dx^\mu}{dx^{\ssc 4}}$ instead of the conventional Einstein
proper time derivative $m \frac{dx^\mu}{d\tau}$. Taking $\kappa\,c$
as the natural momentum unit, we have from Eq.(\ref{pi}) the
momentum $\pi^{\ssc A}$ as essentially nothing more than the
derivative $\frac{dx^{\ssc A}}{ds}$. Taking this to the 6-geometry,
we introduce the natural definition $\pi^{\ssc\mathcal
M}=i\,\frac{dx^{\ssc\mathcal M}}{ds}$. The reason for introducing
the $i$  is the following. The coordinate $x^{\ssc 4}$ has the
opposite metric signature relative to $\tau$ or $x^{\ssc 0}$. Only
with the $i$ we can have the result $\eta_{\ssc\mathcal M\!\mathcal
N} \pi^{\ssc\mathcal M} \pi^{\ssc\mathcal N}=-1$. Now, we can go
along the argument as we have done for $p^{\ssc\mathcal M}$. Write
the conserved quantities as \be \frac{1}{\kappa\,c} J^{\ssc\mathcal
M\!\mathcal N}=x^{\ssc\mathcal M} \pi^{\ssc\mathcal N} -
x^{\ssc\mathcal N} \pi^{\ssc\mathcal M} =i\,\left(z^{\ssc\mathcal M}
\Pi^{\ssc\mathcal N}_{\ssc (\mathcal L)} - z^{\ssc\mathcal N}
\Pi^{\ssc\mathcal M}_{\ssc (\mathcal L)}\right) \;, \ee where \be
\Pi^{\ssc A}_{\ssc (\mathcal L)}= - \frac{1}{\kappa\,c\,\ell}
J^{\ssc{A}5}= i\, G^2  \frac{dz^{\ssc A}}{ds}, \qquad \mbox{and}
\qquad\ \Pi^{\ssc 5}_{\ssc (\mathcal L)}=0 \;. \ee Note that we have
put in the factor ${\kappa\,c}$ to get the unit right. Obviously,
the geodesic equation can also be considered as $\frac{d\Pi^{\!\ssc
A}_{\ssc (\mathcal L)}}{ds} =0$, as $\Pi^{\ssc A}_{\ssc (\mathcal
L)}$ differs from $P^{\ssc A}_{\!\ssc (\mathcal L)}$ only by a
constant factor. The appearance of the $i$ in the above may be taken
as an illustration of the intrinsic quantum nature of the
formulation. In the place of  the energy-momentum constraint of
Eq.(\ref{pi2}), we have instead \be g_{\ssc \!A\!B} \Pi^{\ssc
A}_{\ssc (\mathcal L)} \Pi^{\ssc B}_{\ssc (\mathcal L)}= -G^4 \;,
\ee which reduces to unity at the coordinate origin $z^{\ssc A}=0$.

\section{Summary}
In this paper, we have proposed a simple linear realization of the
{\small \boldmath $SO(1,5)$} symmetry as the Lie-algebraic
description of quantum relativity. The relativity follows from
successive deformations of Einstein special relativity through the
introduction of invariant bounds on energy-momentum and on extended
geometric (``space-time") interval. The invariants are related
respectively to the  Planck scale and the cosmological constant. We
have discussed the logic of our formulation, and plausible physical
interpretations that we consider to be naturally suggested by the
latter. The linear realization has a six-dimensional geometric
description, with the true physics restricted to a dS$_5$
hypersurface embedding the standard four-dimensional Minkowski
space-time. The relativity algebra may be taken as the phase-space
symmetry of the quantum (noncommutative) four-dimensional space-time
with a natural Minkowski limit. We focus mostly on the five or
six-dimensional geometric description with quite unconventional
coordinate(s) (as we have argued) beyond the conventional space-time
ones. There remain several open questions such as a new definition
of energy-momentum in the nonquantum-relativistic limit.  Our
analysis aims at taking a first step in an exploration that may
complement the previous approaches on the subject matter. It
certainly raises some interesting questions that we hope to return
to in future publications.

\acknowledgements Otto Kong would like to thank C.-M. Chen and F.-L.
Lin for helpful discussions, and and members at NCU as well as at
Institute of Physics, Academia Sinica for comments and questions.
This work is partially supported by the research Grant No.
94-2112-M-008-009 from the NSC of Taiwan as well as by the US DOE
Grant No. DE-FG  02-91ER40685.



\begin{thebibliography}{99}
\bibitem{KNncg}
J.~Kowalski-Glikman and S. Nowak, Int. J. Mod. Phys.D {\bf  12}, 299 (2003);
note that the matching of the algebra to our notation here is given by
$\tilde{X}_i \to -\hat{X}_i$ and $\tilde{X}_{\ssc 0} \to \hat{X}_{\ssc 0}$ and
$\ell \to  \frac{1}{\kappa \,c}$ ($\ell$ in the paper is not the same as $\ell$
used here which characterizes the next deformation).
\end{thebibliography}
\end{document}